# A Packet Scheduling Strategy in Sensor Networks with SGMH Protocol

Mary Cherian[1], Dr.Ambedkar Institute Of Technology, Bangalore

T.R.Gopalakrishnan Nair[2], Senior Member IEEE
Dayananda Sagar Institutions, Bangalore, India
thamasha2005@yahoo.com[1]  trgnair@ieee.org[2]

*Abstract*- Data communication in sensor networks can have timing constraints like end to end deadlines. If the deadlines are not met either a catastrophe can happen in hard real time systems or performance deterioration can occur in soft real time systems. In real time sensor networks, the recovery of data through retransmission should be minimized due to the stringent requirements on the worst case time delays. This paper presents the application of Stop and Go Multihop protocol (SGMH) at node level in wireless sensor networks for scheduling and hence to meet the hard real time routing requirements. SGMH is a distributed multihop packet delivery algorithm. The fractions of the total available bandwidth on each channel is assigned to several traffic classes by which the time it takes to traverse each of the hops from the source to the destination is bounded. It is based on the notion of time frames (Tfr). In sensor networks packets can have different delay guarantees. Multiple frame sizes can be assigned for different traffic classes.

.

## I. INTRODUCTION

Sensors are deployed in battle fields, surveillance systems, natural calamity detection systems and environmental monitoring systems. Sensor networks deal with real time environments and their function is data dissemination and data communication. The sensor nodes will generate alarms for certain events. The significance of the data is lost if the alarm message does not reach the base station within the predefined deadline. Therefore it is necessary that the communication should meet the real time constraints [8].

Sensor consists of short range radio transceivers, low power processors and memory with limited capability. Hence they form multihop adhoc networks to communicate among themselves and to the remote base station. Different data streams in the WSN will have different validity intervals and update deadlines which depend on the application. Due to the resource limitation of the individual sensor nodes, the nodes operate in groups. Before sending final information to the base station, the sensors in the local area co-ordinate among themselves to disseminate the data. Reporting of the aggregated data to the base station can be in multiple hops. Group activities require co-ordination and communication among the member nodes. Therefore scheduling of the communication medium is required in meeting the deadlines for real time communication in WSN [4].

Congestion can happen in certain regions in sensor networks. Sudden rise in temperature in certain region of a temperature monitoring system will result in an enormous amount of data communication. Meeting the timing constraints in such overloaded condition is very critical in real time communication. In order to ensure traffic smoothness and to provide loss free communication meeting the deadlines, an admission policy at the edges of the sensor network need to be defined. A bounded delay transmission and service discipline at the intermediate nodes also need to be ensured [2].

The remaining of the paper is organized as follows. Section II gives an overview on the hard and soft real time systems, the real time requirements in WSN and the related work. Section III describes the Stop and Go Multihop Protocol and its performance and section IV gives the conclusion and contributions of the algorithm.

## II. HARD AND SOFT REAL TIME SYSTEMS

Real time systems can be classified as soft or hard depending on the timing constraints. This classification is based on the functional criticality of jobs, usefulness of the late results and the deterministic/probabilistic nature of the constraints [3]. Timing constraint or deadline can be defined as hard if the failure to meet it is considered as a fatal fault. An example could be the failure in meeting the dead line in the data communication in the air traffic control system which will result in a disaster. Tardiness of a job is defined as how late the job completes with respect to its deadline [3]. The result of the hard real time job becomes useless if the tardiness of the job is more than zero. When a task is created with the hard real time constraint by a new application, it is submitted to the scheduler. The scheduler will carry out an acceptance test [3] to see whether the new task can be admitted with its hard real time constraints along with the already running tasks. Such principles can be made use of in WSN to transform them into networks to work with certain real time applications.

### A. Hard Real Time Requirements in WSN and the related work

This section describes the work done in real time transmission of data in WSN. Broadly these solutions may be classified into hard real time and soft real time [6]. Soft real time sensor networks can have two types of messages namely normal messages and real time messages. The communication techniques should handle these messages in different levels of priority. Real time messages should be delivered with highest priority where as in the case of normal messages the miss ratio has to be minimized. The data communication in WSN should meet the end to end deadlines and minimize the packet dead line miss ratio in the worst scenarios.

RAP is a real time communication protocol which applies velocity monotonic scheduling (VMS) [4]. In this case higher priority is assigned for packets which request higher velocity. Compared to the non-prioritized packet scheduling VMS improves the deadline miss ratio of the WSN. Assumption is that each sensor knows its own location by GPS or other services. Velocity is calculated based on the end to end deadlines and the communication distance.

Speed[5] is another real time protocol developed for WSN. Speed and RAP are based on geographic forwarding and are soft real time solutions. In R2TP[10], the packet forwarding is based on the time information. The algorithm tries to achieve reliability by duplicating the packets. The other related work is protocol with Constraint Equivalent delay. The end to end deadline requirement was separated into each link's constraint equivalent delay to enable route discovery process [8].

But all the above algorithms do not consider the priority of the individual packets. The proposed scheduling strategy differentiates the packets based on their priority. Different traffic classes are assigned based on the priority of the packets.

### III. STOP AND GO MULTIHOP PROTOCOL (SGMH)

This is one of the protocols that can be suggested for scheduling the packets and hence meet the hard real time requirements of WSN at the node level. This is a multihop packet delivery algorithm[1].The fractions of the total available bandwidth on each channel is assigned to several traffic classes by which the time it takes to traverse each of the hops from the source to the destination is bounded. The sum of the upper bounds of time on each hop will be the maximum packet delivery time.

### A. Time Frames(Tfr)

Consider the network with n nodes $j = 1...n$ and links $l = 1...L$. Let $C_l$ denote the capacity of the link in bits/second. Let $\Gamma_l$ represent the sum of the propagation delay of the link l, processing and switching time at the receiving end of the link l. $\Gamma_l$ can be considered as a constant for all the links in the network. For every link of the network a Tfr is defined. The Tfr can be of different sizes. Over each link, the corresponding Tfr can be visualized as traveling with the packets from one end of the link to the next end and to the receiving node. The Tfr at the receiving end of the link may be called arriving Tfr of the link and the corresponding Tfr at the transmitting end may be called departing Tfr [2]. The arriving Tfr of the link is delayed by $\Gamma_l$ with respect to the corresponding departing Tfr of the link. The Tfr may be defined as the periodic interval of time that is the logical container within which packets traverse through the network.

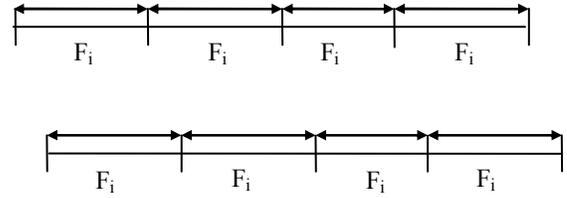

Figure 1. Transmission of Tfr $F_i$

Tfr is not synchronized across the links, but is associated with the network links. There are different Tfr types, each one will define different intervals of time. A virtual Tfr-start signal may be visualized at the input end of each link at the appropriate time. An instance of the Tfr type $F_i$ begins when the previous instance of a $F_i$ ends. Upper Part of Figure 1 indicates the transmission end and lower part indicates the receiver end for a link [1]. Each Tfr type can be assumed to the representing a traffic class [1]. As the algorithm emphasize on bounding the delay at each node in the sensor network, when a packet associated with the Tfr type $F_i$ which may be called TYPE –I packet reaches an intermediate node p on its way to the base station or another node it will be retained by the node p till the beginning of the next instance of the Tfr type $F_i$ and will be transmitted during that Tfr only. It is assumed that it will be possible to bind the number of packets related to each Tfr type so that there will be enough time in each Tfr for every packet associated with that Tfr type to be transmitted[1].

Consider the node P1 which has an input link L and output link $L_1$. Consider the scenario in which the packet of TYPE-I has arrived at the node P1 through the incoming link L. It will be eligible for transmission by the outgoing link $L_1$ during the departing frame of TYPE-I [2]. A link will not stay idle while there is an eligible packet in the queue. According to the class of traffic, the type of the packet is decided and priority will be assigned.

Let there be a TYPE-I packet that is associated with Tfr type $F_i$ which travel from node P1 to node P4, en route the two nodes P2 and P3. In figure 2, the packet has to traverse three hops to reach the destination. Let the propagation time on each link P1 →P2=t12, P2 →P3 = t23, P3 →P4 = t34. Figure 2 depicts the multihop transmission from node P1 to P4.

A connection K is admitted into the network only if its transmission rate $R_k$ is less than or equal to the link capacity of each link from the source to the destination [2]. For a TYPE-I packet the aggregated length of the packet received should be limited to $R_k F_i$ where $F_i$ is the time interval of the Tfr [2]. The aggregate allocated capacities for all the connections in a link should be less than the capacity of the link.

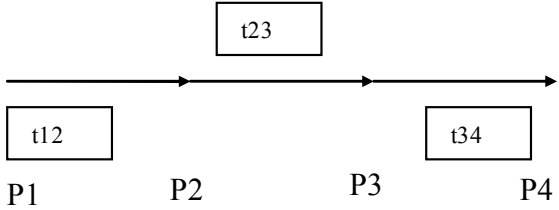

Figure 2. Multihop transmission - P1 to P4

Above criteria must be satisfied while admitting packets to the sensor network [2]. Each connection is set up according to some frame size $F_i$. Here we assume a MAC layer which is collision free. Each packet includes the timing information such as the sending time and the dead line requirements. Since the wireless communication is highly unstable, at each node the elapsed time is computed. If the elapsed time exceeds the dead line requirements the packet is marked as a late packet. The base station will use this information while processing the packet.

*B. Concept of Distribution*

In sensor network where nodes are distributed, each node is autonomous and has the capability to process on its own without a central control. The SGMH protocol suits the environment since it is a distributed algorithm [1]. A TYPE-I packet which reaches a node P will become eligible for transmission only on the following $F_i$. Each sensor node is served in the non preemptive priority order that is, the TYPE-I packet will be having priority over all TYPE –K packets for K<I.

*C. Performance*

The upper bound on the delay incurred by each packet at individual node may be as follows. It has been assumed that the path discovery from the source to the destination has been done through some protocol like R2TP [10]. If we set a certain limit to the loading on the network, a TYPE-I packet will be transmitted within $F_i$ time units once it is eligible for transmission. This will be the upper bound for the delay at each node [2].

Let $D_l$ be the total load on link L from Type-I packets with $C_I$ denoting the total capacity of the link. Let S denote the maximum packet size, and $f_i$ denote the Tfr of TYPE-I packet. A constraint of following form will exist.

$$\sum_{i=j}^{N} D_l \left(1 + \left\lceil \frac{f_j}{f_i} \right\rceil \right) \frac{f_i}{f_j} - D_l \leq \begin{cases} C_I - \dfrac{S}{f_j} ..if\ j = 2 \cdots N \\ C_I \ldots \ldots if \quad j = 1 \end{cases} \quad [1]$$

Here the algorithm provides a way to congestion control by scheduling the packets and meeting the hard real time requirements in WSN by limiting the delay at each node. The time required by a TYPE -I packet to become eligible for transmission is at the most $f_i$ and no packet will be delayed more than 2 $f_i$ at any node [1].

TABLE 1 – Queuing Delay

| Class | Frame Size(ms) | Minimum delay(ms) | Maximum Delay(ms) |
|---|---|---|---|
| TYPE – 1 | $F_1$=1 | 5 | 10 |
| TYPE – 2 | $F_2$=5 | 25 | 50 |
| TYPE - 3 | $F_3$=10 | 50 | 100 |

Table 1 depicts the relation between the frame size and the queuing delay. Since WSN contains real time and best effort traffic [7] we can take three classes of traffic namely TYPE – 1, TYPE – 2 and TYPE – 3.

TYPE-1 has the highest priority and TYPE-3 has the lowest priority. The number of hops between the source and the destination is assumed to be five.

Figure 3 shows the minimum and the maximum queuing delays for different frame sizes obtained from the delay calculations, which indicates the bounded values.

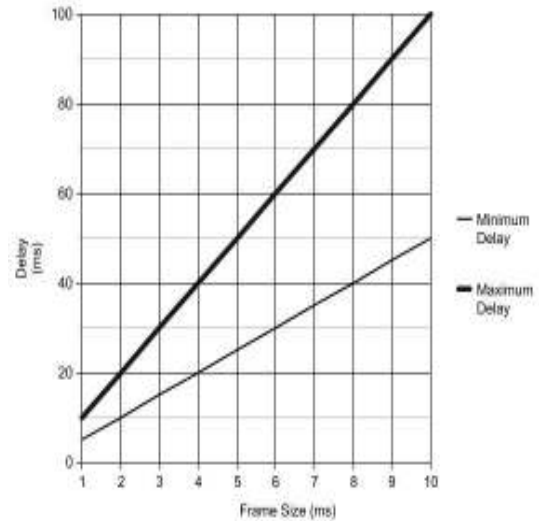

Figure 3. Queuing Delay

### D. Buffer Size

Congestion can happen due to buffer overflow in multihop communication. Shortage of buffer space will result in the loss of packets[9]. In order to reduce the loss of packets, a buffer size at each node will be decided. The buffer size for TYPE-I packets $b_l^i$ is the product of $y_l^i$ which is a constant, $D_l^i$ which is the load on the link l due to TYPE-I packets and $T_i$ is the frame size for TYPE-I packets [2].

$$b_l^i = y_l^i . D_l^i . T_i \quad [2].$$

A certain percentage of the bandwidth will be allocated to each class of traffic [7]. Let the network link capacity is 200 Mb/s and packet size is fixed and the constant $y_l^i$ is 2. With allocating 70%, 20%, 10% of total available bandwidth to TYPE-1, TYPE-2, TYPE-3 packets respectively, Table 2 depicts the buffer size required for each type of traffic.

TABLE 2 – Buffer size allocation

| Class | Frame Size(ms) | Band Width Allocated | Buffer Size(Mb/s) |
|---|---|---|---|
| TYPE – 1 | $F_1$=1 | 70% | 280 |
| TYPE – 2 | $F_2$=5 | 20% | 400 |
| TYPE - 3 | $F_3$=10 | 10% | 400 |

### IV CONCLUSION

This paper has discussed the implementation of SGMH protocol in a sensor network environment. The algorithm provides an efficient way to meet the hard deadlines on packet delivery times by bounding the delay at each node since queuing delay contributes greatly to end to end delay. This results in congestion management in the sensor networks. This protocol prevents packet clustering and provides smoothness to the traffic. Since no assumption is made about the topology of the network the framing strategy will not depend on the topology of the network. The delay time suffered by Type -I packet at the maximum will be 2 $f_i$ to become eligible for transmission at any node. The additive delay on to this will be the packet processing delay at each node and the message propagation delay [1]. The design factors for a specific deadline requirement can be the link capacity and the Tfr size.